\documentclass[manuscript=article, layout=twocolumn]{achemso}
\setkeys{acs}{articletitle = true}
\usepackage[T1]{fontenc}
\usepackage[utf8]{inputenc}
\usepackage{graphicx}
\usepackage{amsmath}
\usepackage{color}
\usepackage{comment}
\usepackage{hyperref}
\usepackage{multirow}
\usepackage{subfig}
\usepackage[normalem]{ulem}

\definecolor{ao}{rgb}{0.0, 0.5, 0.0}

\title{Density matrix renormalization group with dynamical correlation via adiabatic connection}

\author{Pavel Beran}
\affiliation{J. Heyrovsk\'{y} Institute of Physical Chemistry, Academy of Sciences of the Czech \mbox{Republic, v.v.i.}, Dolej\v{s}kova 3, 18223 Prague 8, Czech Republic}
\alsoaffiliation{Faculty of Mathematics and Physics, Charles University, Prague, Czech Republic}
\altaffiliation{Contributed equally.}

\author{Mikuláš Matoušek}
\affiliation{J. Heyrovsk\'{y} Institute of Physical Chemistry, Academy of Sciences of the Czech \mbox{Republic, v.v.i.}, Dolej\v{s}kova 3, 18223 Prague 8, Czech Republic}
\alsoaffiliation{Faculty of Mathematics and Physics, Charles University, Prague, Czech Republic}
\altaffiliation{Contributed equally.}

\author{Micha{\l} Hapka}
\affiliation{Institute of Physics, Lodz University of Technology, \mbox{ul. Wolczanska 219, 90-924 Lodz, Poland}}
\alsoaffiliation{Faculty of Chemistry, University of Warsaw, ul.\ L.\ Pasteura 1, 02-093 Warsaw, Poland}

\author{Katarzyna Pernal}
\email{pernalk@gmail.com}
\affiliation{Institute of Physics, Lodz University of Technology, \mbox{ul. Wolczanska 219, 90-924 Lodz, Poland}}

\author{Libor Veis}
\email{libor.veis@jh-inst.cas.cz}
\affiliation{J. Heyrovsk\'{y} Institute of Physical Chemistry, Academy of Sciences of the Czech \mbox{Republic, v.v.i.}, Dolej\v{s}kova 3, 18223 Prague 8, Czech Republic}

\keywords{Density matrix renormalization group; adiabatic connection, oligoacenes, Fe(II)-porphyrin, 3Fe-4S cluster}

\begin{document}

\begin{abstract}
The quantum chemical version of the density matrix renormalization group (DMRG) method has established itself as one of the methods of choice for calculations of strongly correlated molecular systems. Despite its great ability to capture strong electronic correlation in large active spaces, it is not suitable for computations of dynamical electron correlation. In this work, we present a new approach to the electronic structure problem of strongly correlated molecules, in which DMRG is responsible for a proper description of the strong correlation, whereas dynamical correlation is computed via the recently developed adiabatic connection (AC) technique, which requires only up to two-body active space reduced density matrices. We
report encouraging results of this approach on typical candidates for DMRG computations, namely the $n$-acenes ($n = 2 \rightarrow 7$), Fe(II)-porphyrin, and Fe$_3$S$_4$ cluster.
\end{abstract}

\maketitle

\section{Introduction}
\label{section_introduction}

Strongly correlated molecules (and ions) undoubtedly represent one of the most challenging problems of current quantum chemistry. There exist several approaches to tackle systems for which a single Slater determinant is not a sufficient starting point for a correlation energy treatment. An appealing strategy is to use multireference (MR) formulations of standard computational methods. However, development of robust MR extensions is typically not trivial. The prominent examples are the density functional theory (DFT) \cite{dft_book}, a highly successful single reference approach with a favourable scaling, or the coupled cluster (CC) theory \cite{cizek-original} with CCSD(T) providing a spectroscopic accuracy in single reference cases \cite{gauss_encyclopedia}. Although several MR formulations of DFT \cite{Leininger1997, Gao2016, Chen2017, pdft, pernalrange} and CC \cite{Lyakh2011} have been introduced, none of them is an obvious choice and their development is still an active area of research.

A different approach to proper description of near-degenerate states are the multiconfiguration self-consistent field (MCSCF) methods, among which the complete active space self-consistent field (CASSCF) \cite{casscf} is the most popular. In MCSCF approaches, the dynamical correlation is taken into account by post-SCF methods, such as the complete active space second-order perturbation theory (CASPT2) \cite{caspt2}, the second-order $n$-electron valence state perturbation theory (NEVPT2) \cite{Angeli2001}, or the multireference configuration interaction (MRCI) \cite{mrci}. All of the aforementioned CASSCF-based approaches are limited to small active spaces (less than 20 orbitals) due to the full configuration interaction (FCI) wave function expansion within the active space.

There exists a plethora of highly important molecular systems that require much larger active spaces than CASSCF can provide. Among others, transition metal complexes with multiple transition metal atoms or polycyclic aromatic hydrocarbons (PAHs) belong here. One of the methods with the ability to capture strong correlations in active spaces containing tens of molecular orbitals is the density matrix renormalization group (DMRG). DMRG was originally developed for computations of one-dimensional model systems in solid state physics \cite{White1992, White-1993}. However, since its introduction in quantum chemistry \cite{White1999}, it has established itself as a powerful technique suitable for generic strongly correlated molecules requiring very large active spaces \cite{chan_review, wouters_review, Szalay2015, yanai_review, reiher_perspective}. DMRG computations of biologically relevant complexes with multiple transition metal centers \cite{kurashige_2013, Sharma2014, Li2019, chan_new, Brabec2020} belong to the most advanced quantum chemical applications of DMRG.

Despite the favorable scaling of the DMRG method, it is computationally prohibitive to treat the dynamical electron correlation by including all virtual orbitals into the active space.
Several post-DMRG methods capturing the missing dynamical correlation has been developed.
Probably the most commonly used are the many-body perturbation theories, DMRG-CASPT2 \cite{Kurashige-2011} and DMRG-NEVPT2 \cite{Roemelt2016, Freitag2017}. Their bottleneck however is, that they require the three and four-body reduced density matrix (RDM) elements, which complicates their use in connection with larger active spaces that DMRG itself is suited to handle. A different type of the perturbation theory introduced by Chan and Sharma employs the DMRG-type optimization of the Hylleraas functional \cite{sharma_2014c}. This approach does not need higher-body RDMs, however, it is efficient only if the first-order correction of the wave function can be represented as a matrix product state with a low bond dimension, which may not be the case for large active spaces \cite{reiher_perspective}. On the other hand, a common hurdle of post-DMRG methods employing DFT \cite{srdft,Sharma2019} is that they require a design of new functionals \cite{reiher_perspective}.
Recently, we have explored an alternative direction and developed the single reference coupled cluster (CC) method with singles and doubles tailored by matrix product states \cite{Veis2016}.

In this article, we took a different route and combined DMRG with the recently developed adiabatic connection (AC) theory for multireference wavefunctions \cite{Pernal2018, Pastorczak2018}. This approach can be used in connection with CASSCF-type wave functions \cite{Pastorczak2018} and has the advantage of requiring only the one and two-body active space RDMs. It therefore seems to be an ideal dynamical electron correlation method suitable for large-scale DMRG (or similar methods, where only low-order density matrices are available \cite{Maradzike2020}).

\begin{figure*}[!ht]
  \subfloat[\label{pa_str}]{%
    \includegraphics[width=7cm]{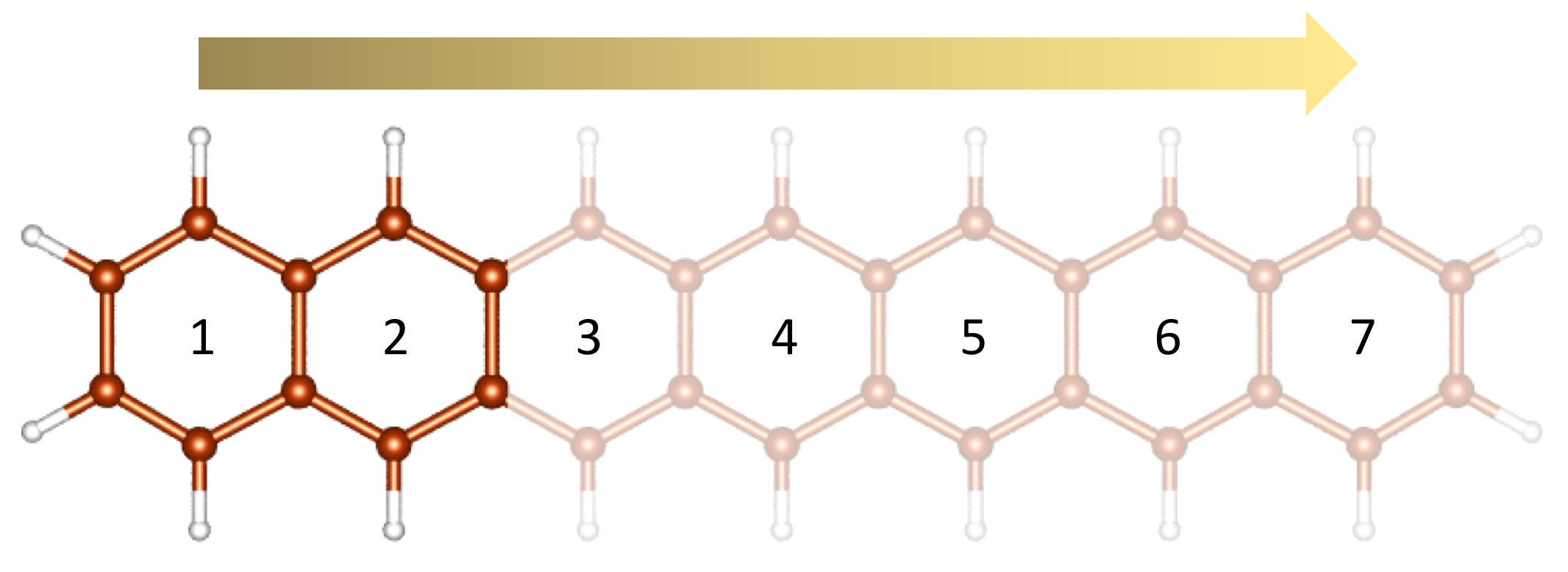}
  }
  \hfill
  \subfloat[\label{porph_str}]{%
    \includegraphics[width=4cm]{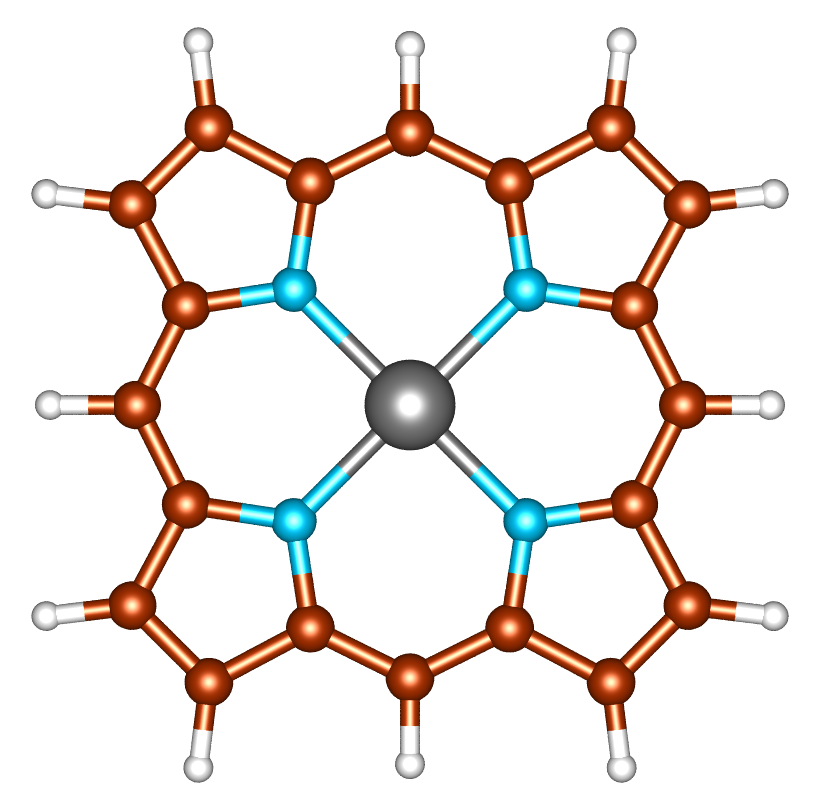}
  }
  \hfill
  \subfloat[\label{fes_str}]{%
    \includegraphics[width=4cm]{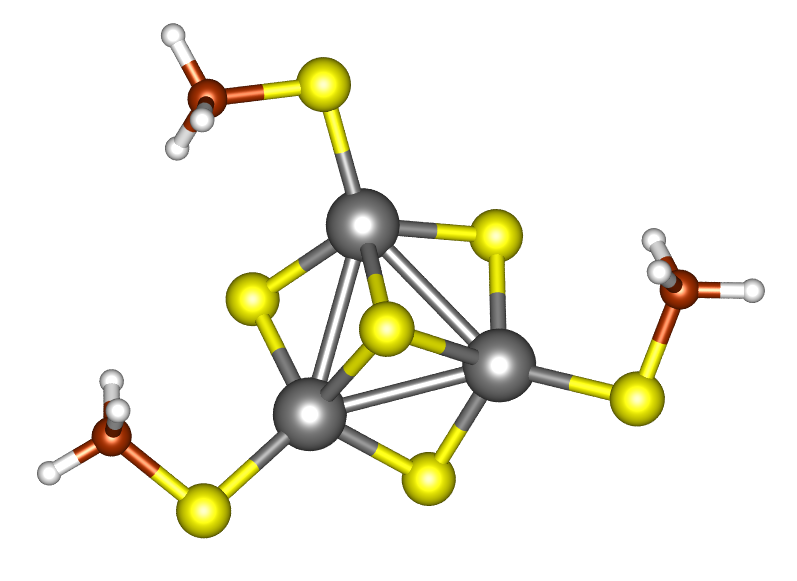}
  }
  \caption{Molecular structures of (a) $n$-acenes series ($n = 2 \rightarrow 7$), (b) Fe(II)-porphyrin, and (c) [Fe$_3$S$_4$(SCH$_3$)$_3$]$^{-2}$ complex. The color codes are as follows: Fe (grey), N (blue), C (brown), S (yellow), and H (white).}
  \label{molecules}
\end{figure*}

In particular, we tested the AC0 approximation \cite{Pernal2018, Pastorczak2018} on top of DMRG and DMRG-SCF reference wave functions on three typical DMRG candidates depicted in Figure \ref{molecules}.
As a first benchmark test, we used the singlet-triplet gaps of $n$-acenes ($n=2 \rightarrow 7$, Figure \ref{pa_str}), for which both experimental values and computational results by several different multireference methods are available. Moreover, the multireference character is varying (increasing) over the $n$-acene series.
Next, we applied the DMRG-SCF-AC0 methodology to the quintet-triplet energy gap problem of Fe(II)-porphyrin (Figure \ref{porph_str}), which has also recently been studied extensively by different multireference computational methods \cite{Vancoillie2011, Phung2016, Zhou2019, Blunt2020} and thus serves as a good (mono-nuclear) transition metal complex benchmark system. As the last system, we took the [Fe$_3$S$_4$(SCH$_3$)$_3$]$^{-2}$ complex (Figure \ref{fes_str}) from our previous study \cite{carlos}, which is characterized by an extremely strong electronic correlation within the active space.

\section{Theory}
\label{section_theory}

In this section, we only briefly remind the basics of both the DMRG methodology, and the adiabatic connection technique. Since we have not modified these methods, but rather connected them, we refer readers interested in more details to comprehensive DMRG reviews \cite{chan_review, wouters_review, Szalay2015, reiher_perspective} or original papers presenting AC/AC0 methods \cite{Pernal2018, Pastorczak2018, Pastorczak2019}.

\subsection{Density matrix renormalization group and matrix product states}
\label{subsection_dmrg}

The DMRG method is a variational procedure, which optimizes the wave function in the so called matrix product state (MPS) form \cite{Schollwock2011}. Let us first express the full configuration interaction (FCI) wave function in the occupation basis representation

\begin{equation}
  | \Psi_{\text{FCI}} \rangle = \sum_{\{\alpha\}} c^{\alpha_1 \alpha_2 \ldots \alpha_n} | \alpha_1 \alpha_2 \cdots \alpha_n \rangle,
\end{equation}

\noindent
where occupation of each orbital reads $\alpha_i \in \{ | 0 \rangle, | \downarrow \rangle, | \uparrow \rangle, | \downarrow \uparrow \rangle \}$ and the expansion coefficients $c^{\alpha_1 \ldots \alpha_n}$ form the FCI tensor. By successive applications of the singular value decomposition (SVD), the FCI tensor can be factorized to the MPS form \cite{Schollwock2011}

\begin{equation}
  \label{mps_factorization}
  c^{\alpha_1 \ldots \alpha_n} = \sum_{i_1 \ldots i_{n-1}} A[1]_{i_1}^{\alpha_1} A[2]_{i_1 i_2}^{\alpha_2} A[3]_{i_2 i_3}^{\alpha_3} \cdots A[n]_{i_{n-1}}^{\alpha_n},
\end{equation}

\noindent
where $\mathbf{A}[j]^{\alpha_j}$ are the MPS matrices (except $j=1$ and $j=n$, which are in this sense vectors) specific to each orbital. In what follows, we will for clarity leave out the $[j]$ notation and write MPS matrices as $\mathbf{A}^{\alpha_j}$. The newly introduced indices $i_j$ by SVD in Eq. \ref{mps_factorization}, which are contracted over, are called virtual indices. If the MPS factorization is exact, the dimensions of the MPS matrices grow in a similar fashion as the size of the original FCI tensor, i.e. exponentially (with an increasing system size). In DMRG, the dimensions of virtual indices are bounded. These dimensions are called bond dimensions and are usually denoted with $M$.

A practical version of DMRG is the two-site algorithm, which provides the wave function in the two-site MPS form

\begin{equation}
  \label{eq:MPS_2site}
  | \Psi_{\text{MPS}} \rangle = \sum_{\{\alpha\}} \mathbf{A}^{\alpha_1} \cdots \mathbf{W}^{\alpha_i \alpha_{i+1}} \cdots \mathbf{A}^{\alpha_n}| \alpha_1 \cdots \alpha_n \rangle,
\end{equation}

\noindent
where for a given pair of adjacent indices $[i, (i+1)]$, $\mathbf{W}$ is a four-index tensor, which corresponds to the eigenfunction of the second-quantized electronic Hamiltonian 

\begin{eqnarray}
  H & = & \sum_{\sigma} \sum_{pq} h_{pq} a_{p_{\sigma}}^{\dagger} a_{q_{\sigma}} + \nonumber \\
    & + & \frac{1}{2} \sum_{\sigma \sigma^{\prime}}\sum_{pqrs} \langle pq | rs \rangle a_{p_{\sigma}}^{\dagger} a_{q_{\sigma^{\prime}}}^{\dagger} a_{s_{\sigma^{\prime}}} a_{r_{\sigma}},
  \label{ham_sec_quant}
\end{eqnarray}

\noindent
expanded in the tensor product space of four
tensor spaces defined on an ordered orbital chain, so called left block ($M_l$ dimensional tensor space), left site (four dimensional tensor space of $i^{\text{th}}$ orbital), right site (four dimensional tensor space of $(i+1)^{\text{th}}$ orbital), and right block ($M_r$ dimensional tensor space).
In Eq. \ref{ham_sec_quant}, $h_{pq}$ and $\langle pq | rs \rangle$ denote standard one and two-electron integrals in the molecular orbital basis, and $\sigma$ and $\sigma^{\prime}$ denote spin.

The MPS matrices $\mathbf{A}$ are obtained by successive application of SVD with truncation on $\mathbf{W}$'s and iterative optimization by going through the ordered orbital chain from left to right and then sweeping back and forth \cite{Szalay2015}. The maximum bond dimension ($M_{\text{max}}$) which is required for a given accuracy,
can be regarded as a function of the level of entanglement in the studied system \cite{Legeza2003}. 
Among others, $M_{\text{max}}$ strongly depends on the order of orbitals along the one-dimensional chain \cite{Barcza2011} as well as their type.\cite{fertitta_2014, Krumnow2016, amaya_2015}.

The DMRG method can be used as a replacement for the exact diagonalization in the CASSCF procedure, which leads to the formulation of a method usually denoted as DMRG-SCF \cite{Zgid-2008c, Ghosh-2008}. Since different elements of the two-body RDMs are collected at different iterations of the DMRG sweep \cite{Zgid-2008b},
the one-site DMRG algorithm, which indeed does not contain any truncation, has to be used for the final computations of the two-body RDMs to assure the same accuracy of all their elements \cite{Zgid-2008c}.
By construction, such two-body RDMs are N-representable \cite{Zgid-2008c} and if obtained from DMRG-SCF procedure with sufficiently large $M$, they are equivalent to RDMs, which would follow from CASSCF calculations in the same active space.

\subsection{Adiabatic connection}
\label{subsection_ac}

The AC theory \cite{Pernal2018} is a general approach to the correlation energy calculation, which can be applied to a broad class of multireference wave functions in the form of group product functions. Since CAS-like wave functions belong to this category, it can be directly applied on top of DMRG or DMRG-SCF. The AC recovers the correlation energy missing in the underlying multireference model and the total electronic energy follows as the sum 
\begin{equation}
 E =E^{\text{DMRG}} +E^{\text{AC}}_{\text{corr}},
\end{equation}
where E is in principle exact in the exact AC formulation. 

The AC formula linearly interpolates between the zeroth-order Hamiltonian $H^{(0)}$ and the exact one, $H$ (Eq. \ref{ham_sec_quant})

\begin{eqnarray}
  H = H^{(0)} + \alpha H^{\prime}, \quad \text{with} \quad H^{\prime} = H - H^{(0)}, \\ \text{and} \quad \alpha:0 \rightarrow 1, \nonumber
\end{eqnarray}

\noindent
where $H^{(0)}$ can be expressed as a sum of group Hamiltonians \cite{Pernal2018}, which for the case of CAS-like reference wave functions comprises only two terms, one corresponding to the doubly occupied (inactive) part and one corresponding to the active orbitals \cite{Pastorczak2018}

\begin{equation}
  H^{(0)} = \sum_I H_I, \quad I \in \{ \text{inactive, active}\}.
\end{equation}

\noindent
The group Hamiltonians have a similar structure to the full Hamiltonian

\begin{eqnarray}
  H_I & = & \sum_{\sigma} \sum_{pq \in I} h_{pq}^{\text{eff}} a_{p_{\sigma}}^{\dagger} a_{q_{\sigma}} + \nonumber \\
   & + & \frac{1}{2} \sum_{\sigma \sigma^{\prime}} \sum_{pqrs \in I} \langle pq | rs \rangle a_{p_{\sigma}}^{\dagger} a_{q_{\sigma^{\prime}}}^{\dagger} a_{s_{\sigma^{\prime}}} a_{r_{\sigma}},   
\end{eqnarray}

\noindent
where $h_{pq}^{\text{eff}}$ denote an effective one-particle Hamiltonian matrix including for a given group $I$ a sum of kinetic and electron-nuclei interaction energy matrix, $h_{pq}$ and a mean-field interaction with electrons in the remaining groups $J$

\begin{equation}
  h_{pq}^{\text{eff}} = h_{pq} + \sum_{J \ne I} \sum_{\sigma} \sum_{r \in J} n_{r_{\sigma}} \langle pr || qr \rangle .
\end{equation}

\noindent
The orbital labels are assumed to correspond to natural orbitals, and $n_{r_{\sigma}}$ refer to occupations of natural spin orbitals.

By exploiting the Hellmann-Feynman theorem and the exact relation between the two-body RDMs and one-body reduced functions, a general AC correlation energy formula can be expressed as

\begin{equation}
  E^{\text{AC}}_{\text{corr}} = \int_0^1 W^{\alpha} d\alpha,
  \label{ac}
\end{equation}

\noindent
where $W^{\alpha}$ contains among others transition density matrix elements between the ground and excited states (for given value of $\alpha$). An important note is that Eq. \ref{ac} already assumes that the one-body RDM stays constant along the AC path. This approximation is justifiable if the CAS reference wave function contains the major part of the static correlation and will be valid only 
if the CAS is big enough. In this respect, AC is especially appealing for connection with DMRG \cite{reiher_perspective}.
In other words, AC correction accounts for (mainly) dynamical correlation, which does not alter the one-body RDM \cite{Pastorczak2018}. 

Extended random phase approximation (ERPA) equations are then employed to approximate the transition density matrices mentioned above. Without going into technical details, which can be found e.g., in Ref. \citenum{Pastorczak2018}, the final generalized eigenvalue problem which has to be solved in order to integrate Eq. \ref{ac} contains only one and two-body active space RDMs.

It was demonstrated numerically in Refs. \cite{Pastorczak2018, Pastorczak2019} that one can avoid the expensive integration in Eq. \ref{ac} by linearized-AC-integrand approximation, $W^\alpha=W^{(0)}+\alpha W^{(1)}_{\alpha=0}$, named AC0, without losing much accuracy. Taking into account that $W^{(0)}=0$ (no correlation at the coupling constant $\alpha=0$), the correlation energy in the AC0 approximation follows as 

\begin{equation}
E^{\text{AC0}}_{\text{corr}} = \frac{1}{2} W^{(1)}_{\alpha=0}.
\label{ac0}
\end{equation}

\noindent
Instead of solving the full ERPA problem as in AC, only
the smaller-sized ERPA equations for specific blocks (active-active, active-inactive, secondary-active, secondary-inactive) at $\alpha = 0$ have to be solved in the case of AC0. This results in an overall scaling $n^2_{\text{sec}} n^4_{\text{act}}$, $n_{\text{sec}} n^5_{\text{act}}$, $n^6_{\text{act}}$, where $n_{\text{act}}$ denotes the number of active orbitals and $n_{\text{sec}}$ denotes the number of secondary (virtual) orbitals.
Numerical accuracy of the AC0 method has been tested on several examples \cite{Pernal2018, Pastorczak2018}, including the complex electronic structure of the tetramethyleneethane diradical \cite{Pastorczak2019} and it was shown that AC0 offers accuracy similar or exceeding that of NEVPT2 \cite{Pastorczak2018}. 
Recently, it has been shown how to amend the AC0 with the contributions from the negative-transition component of the linear response function, in order to achieve good accuracy in description of the singlet excitation energies of organic chromophores \cite{drwal2021}.

In this work, we combine the DMRG(-SCF) and AC0 methods in order to obtain accurate electronic structure of strongly correlated molecules requiring large active spaces, which cannot be treated by CASSCF.
One and two-body RDMs obtained from the DMRG algorithm are employed 
to construct 1-TRDMs 
from solutions of the ERPA equations and subsequently evaluate the AC0 energy correction, Eq.~\eqref{ac0}.
The final electronic energy, denoted as DMRG(-SCF)-AC0, combines DMRG(-SCF) and AC0 correlation energies as

\begin{equation}
 E^{\text{DMRG(-SCF)-AC0}} =E^{\text{DMRG(-SCF)}} +E^{\text{AC0}}_{\text{corr}}.
\end{equation}

The DMRG(-SCF)-AC0 results obtained for oligoacenes and Fe(II)-porphyrin are confronted with their pair density functional theory (PDFT) counterparts.~\cite{pdft} In the PDFT approach, similarly to the AC approximations, only one- and two-body RDMs obtained with a multireference method are required. As opposed to AC, where the correlation energy is directly added to multireference, e.g. CASSCF or \mbox{DMRG(-SCF)}, energy, PDFT combines multireference methods with DFT by retaining one-electron and Hartree energies of CASSCF or DMRG(-SCF) and accounting for the remaining part of electron interaction energy by using exchange-correlation functionals depending on the electron density and the on-top pair density. Recently, it has been shown that AC and PDFT can be seen as limiting cases of the same range-separated  multiconfigurational DFT functional, corresponding to the range-separation parameter approaching $\infty$ and $0$, respectively.~\cite{hapka2020long}

\section{Computational details}
\label{section_computational_details}

As mentioned above, we studied three types of systems which are depicted in Figure \ref{molecules}. 

In the case of oligoacenes, similarly to Ref. \cite{Sharma2019}, we optimized the singlet and triplet state geometries at the UB3LYP/6-31G(d,p) level of theory. The DMRG(-SCF) calculations were then performed in the C-atom p$_\text{z}$ active space employing 6-31G(d,p), 6-31G+(d,p), and cc-pVTZ basis sets. For the production DMRG calculations presented in Tables \ref{polyacenes_st_gaps_vert} and \ref{polyacenes_st_gaps}, we used the pre-defined truncation error (TRE) of $10^{-6}$ together with the dynamical block state selection (DBSS) procedure \cite{legeza_2003a}, which adjusts the actual bond dimensions to fit the desired TRE.
The DMRG-SCF calculations were performed with fixed bond dimensions of $M=1000$. We would like to note that for sufficiently large $M$, the one and two-body RDMs from DMRG-SCF and CASSCF should be numerically identical.

All DMRG calculations presented in this article were initialized with the CI-DEAS procedure \cite{Szalay2015, Legeza2003} and the energy convergence threshold measured between the two subsequent sweeps was set to $10^{-6}$ au. The Fiedler method \cite{Barcza2011} was used for optimization of the orbital ordering.

In the case of Fe(II)-porphyrin, we used the triplet and quintet state PBE0/def2-TZVP geometries taken from Ref. \cite{Phung2016}. For comparison with the SC-NEVPT2 results, we employed the small complete active space (CAS) comprising 8 electrons in 11 orbitals ($3d$ and $4d$ orbitals of Fe together with one $\sigma_{\text{Fe-N}}$ orbital) \cite{Phung2016}. The large (production) active space, CAS(40, 42), is equivalent to the active space used in our previous Fe(II)-porphyrin model study \cite{Antalik2020}, where we have shown that such a large CAS is necessary for quantitatively correct results. This active space contains the $3d$, $4s$, $4p$, and $4d$ orbitals of Fe, the four $\sigma_{\text{Fe-N}}$ orbitals, and the complete $\pi$ space of the conjugated porphyrin ring.
We performed DMRG-SCF(40, 42) calculations with fixed bond dimensions of $M=1000$, followed by a single DMRG run with DBSS and a truncation error fixed to $\text{TRE}=5 \cdot 10^{-5}$ to obtain the one and two-body RDMs for the AC0 energy correction. The aforementioned truncation criterion resulted in DMRG calculations with bond dimensions of up to 13000.
Due to computational limitations of our current in-house AC0 implementation, we employed a hybrid basis, with def2-TZVP on the central Fe atom and def2-SVP on remaining atoms.

For the [Fe$_3$S$_4$(SCH$_3$)$_3$]$^{-2}$ complex,
we used the geometry corresponding to the structure 1E from our recent study \cite{carlos}, which had the lowest energy among the studied Fe$_3$S$_4$ structures.
We used a minimal CAS(15,15), comprising only the Fe $3d$ orbitals. Similarly to the Fe(II)-porphyrin molecule, we used DMRG-SCF with fixed bond dimensions of $M=2000$ (larger than in case of Fe(II)-porphyrin since the open-shell character of Fe-S clusters results in more complex electronic structure requiring higher bond dimensions for its proper description) for the orbital optimization, followed by a single DMRG run with DBSS and $\text{TRE}=10^{-6}$ for the RDMs required by the AC0 energy correction. For comparison, we also performed the SC-NEVPT2 calculations using the same orbitals and a CASCI wave function. We employed the \mbox{cc-pVDZ} basis enlarged by diffuse functions on the Fe and S atoms (aug-cc-pVDZ) to properly describe the anionic character of the complex.

The DFT geometry optimizations of oligoacenes were performed with the Gaussian program \cite{gaussian}, the CASSCF and SC-NEVPT2 calculations were done in Orca \cite{orca}, all DMRG calculations were carried out by means of the MOLMPS program \cite{Brabec2020}, which was for the purposes of DMRG-SCF interfaced to Orca \cite{orca} and the AC0 calculations with our in-house AC program \cite{Pastorczak2018}.

\section{Results and discussion}
\label{section_results}

\subsection{Oligoacenes}
\label{subsection_polyacenes}

\begin{figure*}[!ht]
  \subfloat[\label{st_gaps_fig}]{%
    \includegraphics[width=8cm]{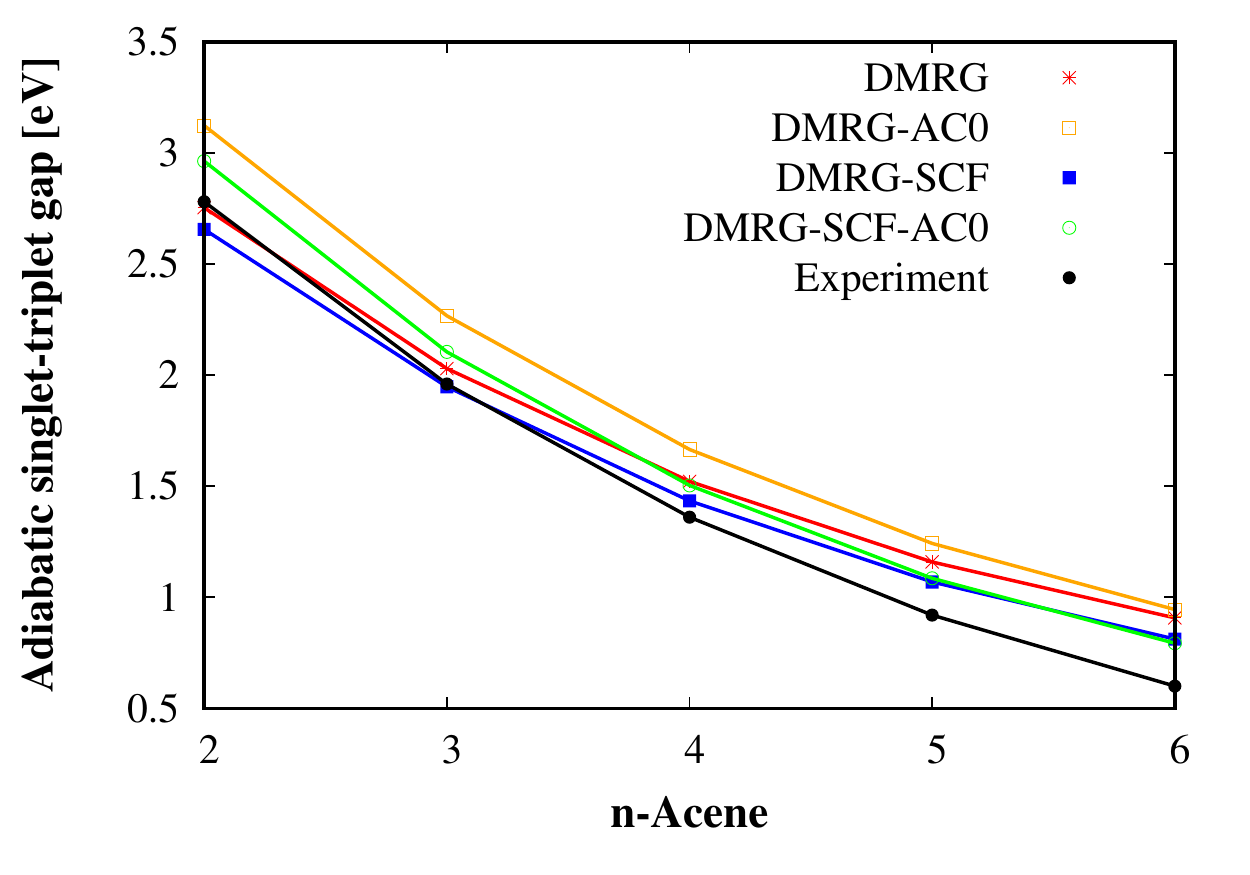}
  }
  \hfill
  \subfloat[\label{occups_fig}]{%
    \includegraphics[width=8cm]{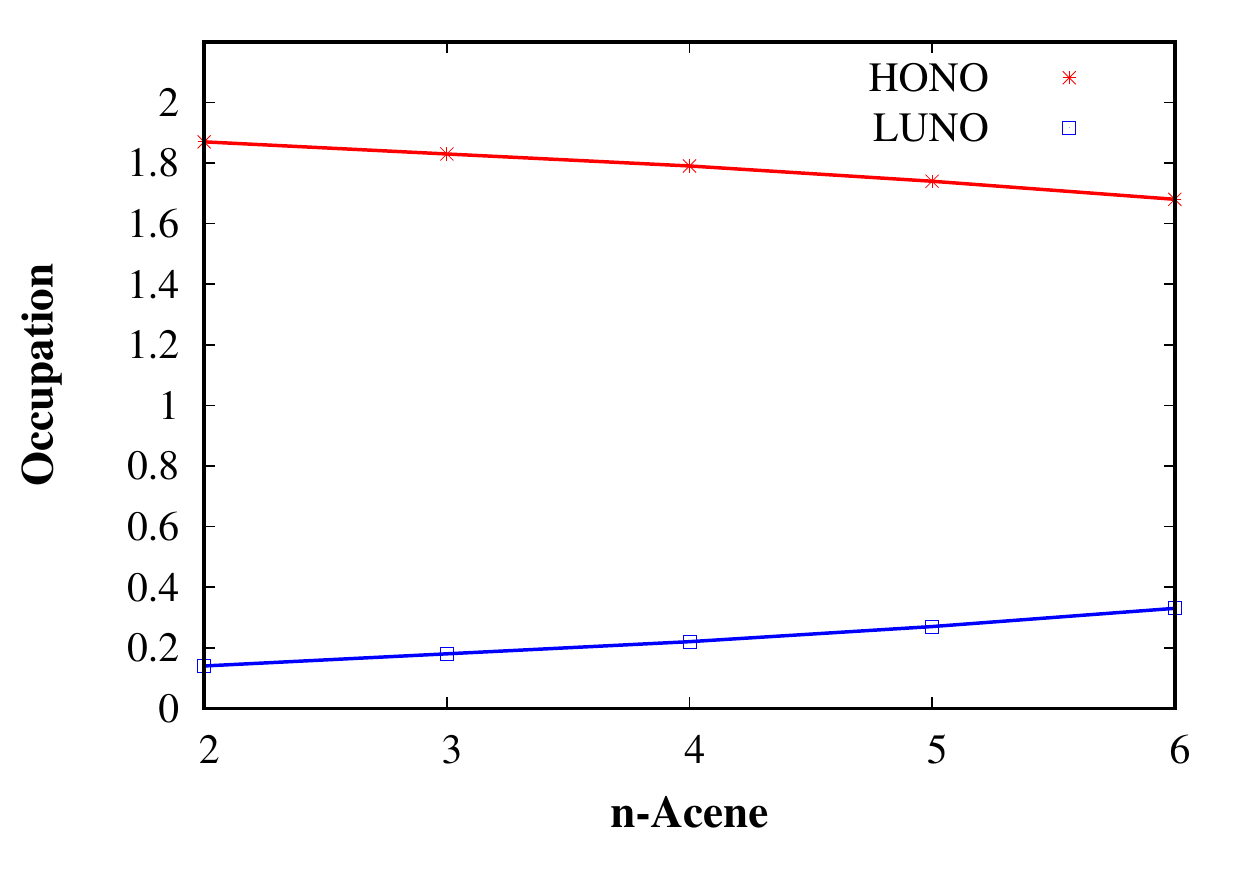}
  }
  \caption{(a) Adiabatic singlet-triplet gaps and (b) DMRG-SCF highest occupied natural orbital (HONO) and lowest unoccupied natural orbital (LUNO) occupations in 6-31G(d,p) basis for oligoacenes. Experimental singlet-triplet gaps are vibrationally corrected.}
  \label{polyacenes_fig}
\end{figure*}

The results on $n$-acenes ($n = 2 \rightarrow 7$) are shown in Tables \ref{tetracene_convergence}, \ref{polyacenes_st_gaps_vert}, \ref{polyacenes_st_gaps}, and Figure \ref{polyacenes_fig}.

Since AC0 has so far been applied mostly in combination with CASSCF wave functions,~\cite{Pernal2018, Pastorczak2018, Pastorczak2019} the first thing to verify when using a DMRG-SCF reference instead is the sensitivity of the AC0 correction to the DMRG accuracy. In the lower part of Table \ref{tetracene_convergence}, we present the dependence of the DMRG-SCF and DMRG-SCF-AC0 vertical singlet-triplet (ST) gaps on the DMRG bond dimensions for tetracene. One can see that similarly to PDFT \cite{Sharma2019}, the DMRG-SCF-AC0 ST gaps are even less sensitive than DMRG-SCF gaps themselves and they are converged already for $M=500$. Moreover, the difference in the ST gap between $M=200$ and $M=2000$ is only about 0.03 eV. This fact justifies using of fixed $M=1000$ for subsequent numerical study of vertical and adiabatic ST gaps in a series from naphthalene to heptacene.

In the upper part of Table \ref{tetracene_convergence}, we show the dependence of the tetracene ST gaps on the DMRG accuracy, when AC0 is applied directly to plain DMRG wave functions. In this case, we used the DBSS approach \cite{legeza_2003a} and set up the truncation error (TRE) a priori. Again, the AC0 energy gaps are less sensitive than DMRG ones and do not change in the range of TRE from $10^{-6}$ to $5\cdot10^{-5}$. For this reason, we used $\text{TRE} = 5\cdot10^{-5}$ in the largest Fe(II)-porphyrin calculations.
  
\begin{table}[!ht]
  \caption{Convergence of DMRG(-SCF-AC0)/6-31G(d,p) vertical singlet-triplet gaps (in eV) with the accuracy of DMRG for tetracene, CAS(18,18).}
  \begin{minipage}{\textwidth}
  \hskip -0.5cm
  \begin{tabular}{l l l l l}
  \hline
  TRE & $5\cdot10^{-5}$ & $10^{-5}$ & $5\cdot10^{-6}$ & $10^{-6}$ \\
  \hline
  DMRG & 1.82 & 1.81 & 1.81 & 1.81 \\
  DMRG-AC0 & 1.94 & 1.94 & 1.94 & 1.94 \\[0.5cm]
  \end{tabular}
  
  \hskip -0.5cm
  \begin{tabular}{l l l l l l}
  \hline
  $M$ & 100 & 200 & 500 & 1000 & 2000 \\
  \hline
  DMRG-SCF & 1.88 & 1.79 & 1.75 & 1.74 & 1.74 \\
  DMRG-SCF-AC0 & 1.88 & 1.82 & 1.79 & 1.79 & 1.79 \\[0.5cm]
  \end{tabular}
  
  \begin{minipage}{\textwidth}
  \begin{footnotesize}
    DMRG-MRCISD+Q$^{a}$ 1.81 , $^a$ Ref. \cite{Zhen2018}.
  \end{footnotesize}
  \end{minipage}
  \end{minipage}
  \label{tetracene_convergence}
\end{table}  
  
\begin{table*}[!ht]
  \caption{Vertical singlet-triplet gaps ($E_{\text{triplet}} - E_{\text{singlet}}$) of $n$-acenes ($n = 2 \rightarrow 7$) in eV. The DMRG(-SCF-AC0) gaps correspond to the 6-31G(d,p) basis.}
  \begin{tabular}{l l l l l l l l l}
    \hline
     n & CAS & DMRG$^{a}$ & DMRG- & DMRG- & DMRG- & DMRG- & CCSD(T)$^{d}$ & DMRG- \\
     & & &  AC0$^{a}$ & SCF$^{b}$ & SCF-AC0$^{b}$ & PDFT$^{c}$ &  & MRCISD+Q$^{e}$ \\
    \hline 
    2 & (10, 10) & 3.09 & 3.52 & 3.05 & 3.36 & 3.35 & 3.30 & 3.43 \\
    3 & (14, 14) & 2.35 & 2.60 & 2.29 & 2.44 & 2.33 & 2.46 & 2.47 \\
    4 & (18, 18) & 1.81 & 1.94 & 1.74 & 1.78 & 1.58 & 1.75 & 1.81 \\       
    5 & (22, 22) & 1.42 & 1.48 & 1.34 & 1.33 & 1.13 & 1.36 & 1.36 \\       
    6 & (26, 26) & 1.14 & 1.15 & 1.05 & 1.01 & 0.79 & 0.99 & 0.98 \\       
    7 & (30, 30) & 0.94 & 0.92 & 0.85 & 0.78 & 0.61 & 0.78 & 0.67 \\       
  \end{tabular}
  \vskip 0.5cm
  \begin{minipage}{\textwidth}
  \begin{footnotesize}
    $^a$ DMRG calculations with DBSS and TRE = $10^{-6}$. 
    $^b$ DMRG calculations with fixed bond dimension $M = 1000$.
    $^c$ Ref. \cite{Sharma2019}.
    $^d$ Ref. \cite{Hajgato2011}.
    $^e$ Ref. \cite{Zhen2018}.
  \end{footnotesize}
  \end{minipage}
  \label{polyacenes_st_gaps_vert}
\end{table*}

\begin{table*}[!ht]
  \caption{Adiabatic singlet-triplet gaps ($E_{\text{triplet}} - E_{\text{singlet}}$) of $n$-acenes ($n = 2 \rightarrow 7$) in eV.}
  \begin{scriptsize}
  \begin{tabular}{l l l l l l l l l l}
    \hline
     n & CAS & DMRG- & DMRG- & DMRG- & DMRG- & CCSD(T)$^{d}$ & DMRG- & ACI-DSRG- & Exp.$^g$ \\
     & & SCF$^{a}/$ & SCF-AC0$^{a}$/ & SCF-AC0$^{a}$/ & PDFT$^{c}$ &  & MRCISD+Q$^{e}$ & MRPT2$^f$ & \\
     & & 6-31G(d,p) & 6-31G(d,p) & cc-pVTZ & & & & & \\
    \hline 
    2 & (10, 10) & 2.66 & 2.96 & 2.74 & 2.91 & 2.85 & 2.71 & 2.70 & 2.78 \cite{Siedbrand1967}, 2.79 \cite{Birks1970} \\
    3 & (14, 14) & 1.95 & 2.10 & 2.11 & 2.00 & 2.09 & 1.81 & 1.87 & 1.95 \cite{Siedbrand1967}, 1.97 \cite{Schiedt1997} \\
    4 & (18, 18) & 1.43 & 1.50 & 1.49 & 1.37 & 1.45 & 1.23 & 1.23 & 1.36 \cite{Siedbrand1967} \\       
    5 & (22, 22) & 1.07 & 1.09 & 1.06$^b$ & 0.98 & 1.10 & 0.92 & 0.78 & 0.92 $\pm$ 0.03 \cite{Burgos1977} \\       
    6 & (26, 26) & 0.81 & 0.79 & 0.76$^b$ & 0.73 & 0.77 & 0.67 & 0.49 & 0.60 $\pm$ 0.05 \cite{Angliker1982} \\       
    7 & (30, 30) & 0.63 & 0.59 & - & 0.62 & 0.58 & 0.48 & 0.33 & - \\       
    \hline
    MAE & & 0.11 & 0.16 & 0.13 & 0.07 & 0.13 & 0.09 & 0.11 &
  \end{tabular}
  \end{scriptsize}
  \vskip 0.5cm
  \begin{minipage}{\textwidth}
  \begin{footnotesize}
    $^a$ DMRG calculations with fixed bond dimension $M = 1000$.
    $^b$ cc-pVTZ(-f) basis on C atoms, cc-pVDZ on H atoms.
    $^c$ Ref. \cite{Sharma2019}.
    $^d$ Ref. \cite{Hajgato2011}.
    $^e$ Ref. \cite{Zhen2018}.
    $^f$ Ref. \cite{Schriber2018}.
    $^g$ Vibrationally corrected experimental values by B3LYP/6-31G(d,p) zero-point energy corrections from Ref. \cite{Sharma2019}.
  \end{footnotesize}
  \end{minipage}
  \label{polyacenes_st_gaps}
\end{table*}

The vertical and adiabatic ST gaps of $n$-acenes for $n = 2 \rightarrow 7$ are presented in Tables \ref{polyacenes_st_gaps_vert} and \ref{polyacenes_st_gaps}, respectively. We will start our discussion with the vertical gaps in order to maximally eliminate the effect of geometries when comparing our results to results of other computational methods. In Table \ref{polyacenes_st_gaps_vert}, one can see that the DMRG-SCF-AC0 vertical ST gaps fit very well with the CCSD(T) \cite{Hajgato2011} as well as the DMRG-MRCISD+Q \cite{Zhen2018} energy gaps. The DMRG-PDFT \cite{Sharma2019} vertical ST gaps are slightly lower. The effect of improvement due to dynamical electron correlation can be clearly seen when looking at the difference between DMRG-SCF and DMRG-SCF-AC0. 

In Table \ref{polyacenes_st_gaps_vert}, we also list the DMRG and DMRG-AC0 ST gaps. The later are noticeably worse (higher) than the DMRG-SCF-AC0 ones, which points out the importance of orbital optimization for the AC0 correction. The mean absolute error (MAE) between DMRG-AC0 and DMRG-SCF-AC0 data is 0.15 eV, which is considerably higher than MAE between DMRG and DMRG-SCF (MAE = 0.07 eV. On the other hand, the difference between DMRG and DMRG-SCF is increasing with $n$ (from 0.04 eV to 0.09 eV), while DMRG-AC0 and DMRG-SCF-AC0 curves are almost parallel, which highlights the fact that AC0 is able to describe systems with varying multireference character in a balanced way.

The adiabatic ST gaps together with experimental values and MAE with respect to them are presented in Table \ref{polyacenes_st_gaps}. In this Table, we show the ST gaps calculated in the 6-31G(d,p) basis as well as in the cc-pVTZ basis (up to $n = 6$). However, due to computational limitations in our current implementation, in case of pentacene and hexacene, the f functions were excluded from the cc-pVTZ basis [denoted as cc-pVTZ(-f)]. Interestingly, the DMRG-SCF method provides ST gaps with lower MAE than DMRG-SCF-AC0. Due to a fortunate cancellation of errors  DMRG-SCF ST gaps of anthracene and tetracene are very close to the experimental values. On the other hand, for larger $n$, the error increases.

In Table \ref{polyacenes_st_gaps}, one can see that the DMRG-SCF-AC0 ST gaps in 6-31(d,p) basis are substantially worse than DMRG-PDFT ones (MAE of 0.16 eV vs 0.07 eV). However, the PDFT results used a larger basis 6-31+G(d,p), which contains diffuse functions. Unfortunately, we were able to converge the DMRG-SCF in the 6-31G+(d,p) basis only for $n=2,3,4$ due to linear dependencies in the basis and got the following DMRG-SCF-AC0 adiabatic ST gaps: 2.91 eV, 1.99 eV, and 1.42 eV. It is therefore apparent that the diffuse functions in the basis might be important especially for larger acenes which have a stronger biradical character and unpaired electrons localized on the edges of a molecule \cite{Hachmann2007}. The improvement of MAE when going to cc-pVTZ is clear. As can be observed in Table \ref{polyacenes_st_gaps}, DMRG-SCF-AC0 in the cc-pVTZ basis achieves the accuracy comparable to CCSD(T) and ACI-DSRG-MRPT2.

In Figure \ref{st_gaps_fig}, we graphically compare the DMRG, DMRG-AC0, DMRG-SCF, and DMRG-SCF-AC0 adiabatic ST gaps in the smaller 6-31G(d,p) basis with the experimental values. One can see that due to some cancellation of errors, DMRG and DMRG-SCF gaps for small-$n$-acenes are in good agreement with the experiment, which is however changing with increasing $n$. In contrast, the AC0 methods provide balanced accuracy for the whole series with varying multireference character (see Figure \ref{occups_fig} for natural orbitals occupations). The DMRG-SCF-AC0 is obviously superior to DMRG-AC0.

\subsection{Iron-porphyrin}

\begin{table}[!ht]
  \caption{Quintet-triplet energy gaps ($E_{^3\text{A}_{\text{2g}}} - E_{^5\text{A}_{\text{1g}}}$) of Fe(II)-porphyrin in eV.}
  \begin{tabular}{l l l}
  \hline
  CAS & Method &  Q-T gap \\
  \hline
  (8, 11) & CASSCF & ~0.82 \\
  (8, 11) & SC-NEVPT2 & ~0.62 \\
  (8, 11) & CASSCF-AC0 & ~0.54 \\
  (8, 11) & CASPT2$^a$ \cite{Vancoillie2011} & ~0.21 \\
  (32, 29) & HCI-SCF$^b$ \cite{Blunt2020} & ~0.85 \\  
  (32, 29) & SC-NEVPT2(s)$^b$ \cite{Blunt2020} & ~0.33 \\
  (34, 35) & DMRG-SCF$^a$ \cite{Zhou2019} & ~0.78 \\
  (34, 35) & DMRG-PDFT:tPBE$^a$ \cite{Zhou2019} & -0.42 \\
  (34, 35) & DMRG-PDFT:ftPBE$^a$ \cite{Zhou2019} & -0.68 \\
  (40, 42) & DMRG-SCF & ~0.44 \\
  (40, 42) & DMRG-SCF-AC0 & ~0.31 \\
  \end{tabular}
  \vskip 0.5cm
  \begin{minipage}{\textwidth}
  \begin{footnotesize}
    $^a$ ANO-RCC basis contracted to 7s6p5d3f2g1h/4s3p2d1f/ \\ 3s1p for Fe/C,N/H.
    $^b$ geometry optimized for the triplet \\ state.
  \end{footnotesize}
  \end{minipage}
  \label{porphyrin}
\end{table}

The computed quintet-triplet (Q-T) energy gaps listed in Table \ref{porphyrin} represent the difference between the two lowest spin states, which at our level of theory (DMRG-SCF as well as DMRG-SCF-AC0) correspond to $^5$A$_{\text{1g}}$ and $^3$A$_{\text{2g}}$ states with electron configurations (d$_{\text{x}^2 - \text{y}^2}$)$^1$(d$_{\text{z}^2}$)$^2$(d$_{\text{xz}}$)$^1$(d$_{\text{yz}}$)$^1$ (d$_{\text{xy}}$)$^1$ and (d$_{\text{x}^2 - \text{y}^2}$)$^2$(d$_{\text{z}^2}$)$^2$(d$_{\text{xz}}$)$^1$(d$_{\text{yz}}$)$^1$, respectively.
In Table \ref{porphyrin}, we present the results in the small CAS(8, 11), in order to compare with the NEVPT2 and CASPT2 energy gaps and some of the recent large-active-space results, which we compare to our DMRG-SCF(40, 42)-AC0 Q-T gap.
  
First of all, one can see that in the case of CAS(8, 11), the CASSCF-AC0 energy gap agrees well with the SC-NEVPT2 one (difference of 0.08 eV). The difference between these two and CASPT2 is more pronounced and can be partly assigned to a larger basis being used in the CASPT2 study (ANO-RCC contracted to triple-$\zeta$ basis set quality) \cite{Vancoillie2011}, since smaller bases are known to be biased toward the high spin state \cite{LiManni2019}. Adding dynamical correlation on top of the CASSCF reference clearly decreases the Q-T gap \cite{LiManni2018}.
The Q-T gap is also decreased (triplet state becoming more stabilized) with enlarging the active space, going from 0.82 eV [CAS(8, 11)] down to 0.44 eV [CAS(40, 42)]. The AC0 energy correction further decreases this gap to 0.31 eV, which is in a good agreement with the recent stochastic SC-NEVPT2 study \cite{Blunt2020} yielding 0.33 eV. One must however take into account a slightly different active space and more importantly the vertical Q-T gap on the triplet state geometry in Ref. \cite{Blunt2020}.
  
The most striking discrepancy is between DMRG-SCF-AC0 and DMRG-PDFT, where the same geometries were used. One would expect that when using a large-enough active space, the CASSCF method should capture also a large amount of dynamical electron correlation and when increasing the active space, the effect of the missing (out-of-CAS) dynamical electron correlation will subside (eventually going to zero with CAS being the full orbital space). This assumption is in agreement with our recent DMRG tailored CCSD(T) study of the Fe(II)-porphyrin model \cite{Antalik2020} and our AC0 results in Table \ref{porphyrin}.
However, in the case of DMRG-PDFT the effect of the missing dynamical correlation for CAS(34, 35) is about 1.2 eV for the tPBE functional and even almost 1.5 eV (33.7 kcal/mol) for ftPBE. We think that this might be caused by too small bond dimensions in DMRG calculations\footnote{In Ref. \cite{Zhou2019}, the convergence with respect to $M$ was verified on a very small CAS(8, 11) and the largest $M=300$ (even though the DMRG was spin-adapted) is not enough for accurate calculations in 35 active orbitals where the Hilbert space is enormously large. Moreover, Fe(II)-porphyrin is not a (quasi-)linear molecule where small bond dimensions are justifiable.}. Another reason may be an empirical character of the PDFT method and the fact that the correlation energy measured as a difference between PDFT and the corresponding CAS energy does not tend to $0$ with the increase of the active space, which may result in erratic behavior if too large active space if employed.\cite{sharma2018active,hapka2020long} This is in contrast to the AC and perturbation methods.

We would like to note that we do not, on purpose, list the available CASSCF Q-T energy gaps for even a larger 44-orbital active space [CAS(44, 44)] in Table \ref{porphyrin}. The reason is that the two recent studies \cite{Smith2017, Levine2020}, which both used selected CI methods as FCI solvers in CASSCF, the same triplet state optimized geometry and cc-pVDZ basis show a discrepancy on the order of 0.9 eV ($\approx$ 21 kcal/mol \cite{Lee2020}, HCI-SCF: -0.09 eV \cite{Smith2017}, ASCI-SCF: 0.84 eV \cite{Levine2020}) and further verification is probably necessary to access their accuracy.
  
Direct comparison of Q-T gaps from Table \ref{porphyrin} with experimental values is unfortunately not possible,
since the Fe(II)-porphyrin molecule is not stable unsubstituted. Even though the existing experimental studies on four-coordinated Fe(II) embedded in substituted porphyrin systems \cite{Spartalian1979,Evangelisti2002,Filoti2006,Bartolom2010,Gruyters2012, Kitagawa1979,Collman1975,Mispelter1980, Sams1974} mostly predict the triplet ground state, they were performed either in the crystal phase or polar solvent, far from the gas phase conditions of computational studies. On the other hand, the recent M\"ossbauer spectroscopy study of Fe(II)-phtalocyanine \cite{Nachtigallova2018} have unambiguously indicated the triplet ground state in the crystalline form and dissolved in dimethylformamide, and the quintet when dissolved in the less polar monochlorobenzene (better resemblance of the gas phase conditions of computational studies). These findings were confirmed by our DMRG tailored CCSD(T) [TCCSD(T)] study of the Fe(II)-porphyrin model \cite{Antalik2020}, in which we have pointed out the crucial importance of the geometry (especially the Fe-N bond length) on the spin state ordering of Fe(II)-porphyrin. 
The vertical DLPNO-TCCSD(T)/def2-TZVP Q-T gap of the smaller Fe(II)-porphyrin model in the active space equivalent to the aforementioned CAS(40, 42) on the B97-D3/TZVPP optimized triplet state geometry corresponds to 0.34 eV (in a very good agreement with DMRG-SCF-AC0 on the full Fe(II)-porphyrin) and the adiabatic Q-T gap to 0.69 eV \cite{Antalik2020}. One must however take into account that the larger value of the adiabatic Q-T gap might be caused by larger flexibility of the model system (when compared to the full Fe(II)-porphyrin molecule) and consequently longer Fe-N bond in the quintet state. In summary, the experimental results on Fe(II)-phtalocyanine as well as state-of-the-art multireference computations on simple model of Fe(II)-porphyrin both indicate that the adiabatic Q-T gap of the Fe(II)-porphyrin molecule should be positive, which is in agreement with our DMRG-SCF-AC0 results.

The auxiliary field quantum Monte Carlo (AFQMC) study \cite{Lee2020} on the triplet state optimized geometry predicts the Q-T gap to be -0.29 eV (in cc-pVTZ basis), which confirms that DMRG-PDFT probably overstabilizes the triplet state (note that DMRG-PDFT results in Table \ref{porphyrin} correspond to the adiabatic Q-T gaps, which should be smaller than the vertical ones). 

\subsection{Iron-sulfur complex}

The computed absolute energies of the four lowest spin states ($2S + 1 = 2,4,6,8$) of the [Fe$_3$S$_4$(SCH$_3$)$_3$]$^{-2}$ complex are presented in Table \ref{fes}. The DMRG energies correspond to DMRG calculations with TRE=$10^{-6}$, which followed the DMRG-SCF orbital optimization and provided the RDMs for AC0 corrections. 

\begin{table}[!ht]
  \caption{DMRG, DMRG-SCF-AC0 and NEVPT2 absolute energies ($E - 6689.0$ Ha) of various spin states of the [Fe$_3$S$_4$(SCH$_3$)$_3$]$^{-2}$ complex in CAS(15, 15). DMRG energies correspond to DMRG(TRE=$10^{-6}$) on top of DMRG-SCF with $M=2000$.}
  \small
  \begin{tabular}{l l l l}
  \hline
  $2S+1$ & DMRG & DMRG-SCF-AC0 & NEVPT2\\
  \hline
  2 & -0.366911 & -3.116489 & -3.026214\\
  4 & -0.366855 & -3.116457 & -3.026785\\
  6 & -0.366898 & -3.116548 & -3.026148\\
  8 & -0.366213 & -3.113951 & -3.024493\\
  \end{tabular}
  \label{fes}
\end{table}

One can see that all spin states are very close in energy at the DMRG level of theory (submilliHartree differences).
Such quasi-degeneracy of spin states has been already reported for similar Fe$_2$S$_2$ and Fe$_4$S$_4$ clusters in Ref. \citenum{sharma_2014b}, where it was shown that \textit{ab initio} DMRG energy spectra of these clusters differ from those predicted by the simplistic Heisenberg double-exchange model. 
The dense energy spectrum is a result of the fact that Fe $d$ orbitals are quasi-degenerate.
Moreover as was shown for Fe$_4$S$_4$ clusters \cite{carlos},
the correlation energy of Fe $d$ orbitals is small,
when compared to S $p$ orbitals, which are responsible for double-exchange stabilization.
In order to compare with the NEVPT2 energies, 
our CAS is minimal and thus does not contain any sulfur orbitals. 
Nevertheless, the recent study \cite{LiManni2021} has shown that this minimal CAS can reasonably predict the energy gaps of iron-sulfur clusters.

Taking into account the very dense DMRG energy spectrum discussed above, one can expect that dynamical electron correlation might play a decisive role in the final spin state ordering, which is confirmed by our DMRG-SCF-AC0 results. In fact, the sextet becomes the ground state, nevertheless the energy gap between doublet and sextet is only about 13 cm$^{-1}$, clearly below the expected precision of the method. In fact, the NEVPT2 theory predicts a quartet ground state, with a similarly negligible spin gap. A similar spin state reordering when going from RASCI to RASCI-PDFT correlated treatment was reported also for Fe$_2$S$_2$ clusters \cite{Presti2019}. Another observation in Table \ref{fes} is that the octet spin state is at the AC0 level of theory separated from the remaining spin states by more than 2.5 mHa, 3.6 times more than at the DMRG level, which can be also seen in the NEVPT2 results. Similar separation can be seen also for Fe$_2$S$_2$ clusters and PDFT theory \cite{Presti2019}. 

Overall, while comparing the AC0 and NEVPT2 results gives us a roughly 90 mHa difference in the absolute energy, the difference does not change significantly over different spin states.
Both AC0 as well as NEVPT2 predict the first three spin states ($2S+1=2,4,6$) to be very close in energy (within 1 mHa) and the fourth one ($2S+1=8$) to be separated by roughly 2 mHa. Taking into account the complexity of the electronic structure of iron-sulfur clusters, such an agreement between AC0 and NEVPT2 is very good, indeed.
Unfortunately, we do not have any experimental data against which we could validate our DMRG-SCF-AC0 results. The current [Fe$_3$S$_4$(SCH$_3$)$_3$]$^{-2}$ example, which we have picked up from our very recent DMRG-SCF study of Fe$_3$S$_3$ and Fe$_4$S$_4$ clusters \cite{carlos} serves mainly as a proof-of-principle benchmark. A detailed DMRG-SCF-AC0 study of the Fe$_4$S$_4$ clusters is a subject of following studies.

Finally, we would like to demonstrate the computational advantage of AC0 over NEVPT2 on CPU timings. In the case of the doublet state of the [Fe$_3$S$_4$(SCH$_3$)$_3$]$^{-2}$ complex, the AC0 correction took roughly 12 hours on a 24-core Intel Xeon machine with the only parallelization at the BLAS level, while NEVPT2 correction took more than 28 days of computation on 8 cores (we could not fully use all cores due to memory demands). For completeness, one DMRG calculation took approximately 10 hours.


\section{Conclusions}
\label{section_conclusionns}

In this article, we have presented a new computational approach to the electronic structure problem of strongly correlated molecular systems, which combines the DMRG and AC methodologies. The former method is responsible for proper description of strong correlations within the active space, while the latter corrects for the missing dynamical electron correlation by means of the computationally favourable AC0 correction. The AC0 correction itself \cite{Pernal2018, Pastorczak2018} requires only the one- and two-body active space RDMs, which makes it very appealing for connections with large-active-space methods like DMRG.

On the example of oligoacenes, we have shown that the DMRG-SCF-AC0 method provides very accurate ST gaps. Unfortunately there is strong evidence that basis sets with diffuse functions would further improve the results, however we were not able to use them systematically for all studied systems due to numerical issues. We have also shown that the simple DMRG-AC0 method can describe systems with varying multireference character in a well-balanced way, however, for quantitative results, orbital optimization is necessary.

Additional examples comprised transition metal complexes, namely Fe(II)-porphyrin and the [Fe$_3$S$_4$(SCH$_3$)$_3$]$^{-2}$ complex. Our DMRG-SCF-AC0 results on the spin state ordering of Fe(II)-porphyrin agree well with the available computational data as well as the recent experimental results on Fe(II)-phtalocyanine \cite{Nachtigallova2018}. On the example of the [Fe$_3$S$_4$(SCH$_3$)$_3$]$^{-2}$ complex, we have shown
a good agreement between AC0 and NEVPT2, which makes the DMRG-SCF-AC0 methodology a suitable candidate for electronic structure studies of more complex iron-sulfur clusters.

\section*{Acknowledgment}
We would like to thank Dr. J. Pittner, who initiated our collaboration, for insightful discussions.

This work has been supported by the the Czech Science Foundation (grant no. \mbox{18-18940Y}),
the Charles University in Prague (grant no. CZ.02.2.69\slash0.0\slash0.0\slash19\_073\slash0016935),
the Center for Scalable and Predictive methods for Excitation and Correlated phenomena (SPEC), which is funded by the U.S. Department of Energy (DOE), Office of Science, Office of Basic Energy Sciences, the Division of Chemical Sciences, Geosciences, and Biosciences,
and by the National Science Center of Poland (grant no. 2019/35/B/ST4/01310). \color{black}

Most of the computations were carried out on the Salomon and Barbora supercomputers in Ostrava, the authors would therefore like to acknowledge the support by the Czech Ministry of Education, Youth and Sports from the Large Infrastructures for Research, Experimental Development and Innovations
project “IT4Innovations National Supercomputing Center—LM2015070.”
\bibliography{references}

\end{document}